# On the universality of the critical scaling exponents during sol-gel transition

Khushboo Suman and Yogesh M. Joshi*

Department of Chemical Engineering, Indian Institute of Technology Kanpur 208016, India

* Corresponding Author, E-mail: joshi@iitk.ac.in,

Tel.: +91 512 2597993, Fax: +91 512 2590104

Abstract:

The evolution of viscoelastic properties near the sol–gel transition is studied by performing oscillatory rheological measurements on two different types of systems: a colloidal dispersion and a thermo-responsive polymer solution under isothermal and non-isothermal conditions. While undergoing sol–gel transition, both the systems pass through a critical point. An approach to the critical point is characterized in terms of divergence of zero shear viscosity and the subsequent appearance of the low frequency modulus. In the vicinity of the critical gel state, both the viscosity and the modulus show a power-law dependence on relative distance from the critical point. Interestingly, the longest relaxation time has been observed to diverge symmetrically on both the sides of the critical point and also shows a power-law dependence on relative distance from the critical point. The critical (power-law) exponents of the zero-shear viscosity and modulus are observed to be related to the exponents of the longest relaxation time by the hyper scaling laws. The dynamic critical exponent has also been calculated from the growth of the dynamic moduli. Remarkably, the critical relaxation exponent and dynamic critical exponent predicted from the scaling laws precisely agree with the experimental values from the isothermal as well as non-isothermal experiments. The associated critical exponents show remarkable internal consistency and universality for different kinds of systems undergoing the sol-gel transition.

## I. Introduction:

The transformation of a polymeric or a colloidal sol with liquid-like properties to a gel state comprised of a percolated space spanning network, either spontaneously or due to change in temperature, is a very common phenomenon in many industrial materials [1-3]. The spontaneous sol-gel transition is observed in a chemical crosslinking process in unsaturated polymeric systems triggered by a chemical curing agent [4] as well as in colloidal dispersions that form physical bonds leading to a three-dimensional network [5]. The sol-gel transition induced by a change in temperature, which many times is reversible in nature, occurs due to alteration of the interparticle/molecular interactions. At the cusp of the sol-gel transition, most of the systems pass through a unique state, known as the critical gel state at which the space spanning percolated network structure is the weakest [6,7]. This state shows distinctive characteristic features wherein the rheological response functions such as dynamic moduli, relaxation modulus, and compliance show a power-law dependence on their respective independent variables, frequency or time [6]. Over past few decades, significant amount of work has been carried out on polymeric materials undergoing chemical crosslinking, which show that the zero shear viscosity in the sol state and the equilibrium modulus in the gel state also show power-law dependence on the distance to critical gel point in terms of the extent of a chemical reaction [8-11]. The corresponding power-law exponents or the critical exponents have been observed to show exclusive interdependence rendering greater insights into the nature of the sol-gel transition [8]. The interdependency amongst various critical exponents is expressed by the hyperscaling relations. While the work on polymeric systems undergoing chemical crosslinking reaction is rich with studies on critical exponents [4,8,10,12,13], the literature on colloidal and polymeric systems undergoing sol-gel transition through physical interactions, either spontaneous or temperature induced, is sparse. The objective of the present work is to study the systems that undergo sol-gel transition through physical interactions and to investigate the universality of the critical exponents by altering the trigger for such a transition.

The linear viscoelastic behavior of a polymeric material undergoing a crosslinking reaction has been proposed and experimentally verified by the seminal works of Winter and coworkers [14,15]. Winter and Chambon [14] report that at the critical gel state, where the system forms the weakest space spanning percolated network, the elastic $(G')$ and viscous $(G'')$ moduli exhibit identical power-law dependence on frequency $(\omega)$ given by:

$$G' = G'' \cot(n\pi/2) = \frac{\pi S}{2\Gamma(n)\sin(n\pi/2)} \omega^n, \qquad (1)$$

where $n$ is the critical exponent restricted between 0 and 1, $S$ is the gel network strength and $\Gamma(n)$ is the Euler gamma function of $n$. The identical power-law dependence of $G'$ and $G''$ on $\omega$ leads to a distinctive feature of the critical gel state. As a result, the loss tangent, given by $\tan\delta = \tan(n\pi/2) = G''/G'$, becomes independent of $\omega$ suggesting it to be independent of the probing timescale [14,15]. In addition, the stress relaxation modulus $G(t)$ of a critical gel is observed to demonstrate the power-law decay with respect to time given by [16]:

$$G(t) = St^{-n}. \qquad (2)$$

Furthermore, the creep compliance $J(t)$ of the critical gel is also observed to increase with time in a power-law fashion given by [16]:

$$J(t) = \frac{(1-n)}{S\,\Gamma(2-n)\Gamma(1+n)} t^n. \qquad (3)$$

The above equations (1), (2) and (3) relating the rheological response functions to their respective independent variables adhere to the principles of linear viscoelasticity and are internally consistent with respect to the convolution relation $t = \int_0^t G(a)J(t-a)da$ and the Fourier transform relation: $G^*(\omega) = i\omega \int_0^\infty G(t)e^{-i\omega t}dt$. The power-law rheology demonstrated by a material at the critical gel state results from a microstructure that is fractal in nature [17]. The resulting self-similar morphology over a wide range of length-scales gives rise to hierarchical relaxation timescales. The power-law dependence of $G(t)$ given by equation (2) leads to a continuous relaxation time spectrum given by [6]:

$$H(\tau) = \frac{S}{\Gamma(n)} \tau^{-n}. \tag{4}$$

Such dependence signifies the scale-invariant relaxation behavior of the critical gel [6]. The above discussion suggests that the rheological behavior exhibited by a critical gel is governed by the critical power-law exponent $n$ in all the material properties. Interestingly, the knowledge of $n$ does not just provide information about the rheological behavior of a critical gel but also can be related to the fractal dimension $(f_d)$ of the associated hierarchical network structure [17].

In a polymeric system undergoing a chemical crosslinking reaction, it is common to monitor the degree of crosslinking defined as the fraction of (crosslinkable) monomeric units that participate in the crosslinks and is expressed as $p$. The degree of crosslinking is equal to 0 and 1 respectively at the beginning and the end of crosslinking reaction. The degree of crosslinking at the critical gel state is defined as $p_c$. Consequently, the progress of a chemical crosslinking process can be monitored by expressing normalized distance from the critical gel state $(\Delta \tilde{p})$ in both the pre-critical gel state (pre-gel) as well as post-critical gel (post-gel) state as: $\Delta \tilde{p} = |p - p_c|/p_c = \Delta p/p_c$ [18]. In the vicinity of the critical gel point, a polymeric system undergoing a chemical crosslinking reaction demonstrates a critical behavior, wherein the size of the largest cluster $(\xi)$ in a pre-gel state grows as [11]:

$$\xi \sim \Delta\tilde{p}^{-\alpha}. \tag{5}$$

Equation (5) suggests that $\xi$ diverges and spans the space as the critical point is approached marking a transition from a sol state to a gel state. Consequently, as a critical gel state is achieved, the molecular weight of the polymer diverges and no viscous flow is possible in a permanent percolated network formed by the covalent bonds. This necessitates the zero shear viscosity also to diverge as [18,19]:

$$\eta_0 \sim \Delta\tilde{p}^{-s}. \tag{6}$$

The corresponding dominant relaxation time of a polymer system undergoing a crosslinking reaction in the sol state diverges as [20]:

$$\tau_{\max,S} \sim \Delta\tilde{p}^{-\nu_S}, \tag{7}$$

while approaching the critical gel state. It is important to note that the scaling law given by Eqs. (6) and (7) are related to the pregel state. The equilibrium modulus ($G_e$, equivalent to elastic modulus in a limit of zero frequency) is theoretically zero in the sol state and is expected to become nonzero and grow only beyond the critical gel point. However, it remains below the detection limit in the immediate vicinity of the critical state but eventually grows as the crosslink density increases according to [18,19]:

$$G_e \sim \Delta\tilde{p}^{z}. \tag{8}$$

Furthermore, in the post gel state, the characteristic relaxation time associated with the finite relaxation modes (the relaxable modes, whose population decreases as the crosslinking reaction progresses) evolves as [20]:

$$\tau_{\max,G} \sim \Delta\tilde{p}^{-\nu_G}. \tag{9}$$

It is interesting to note that the scaling laws given by Eqns. (6) and (8) are based on the assumption that an analogy exists between the mechanical and dielectric properties of a fractal gel system [11,19].

For a chemically crosslinking system approaching the critical gel state, Scanlan and Winter [8] experimentally observed that the rate of change of the complex modulus with respect to $p$ while approaching the critical gel state from the sol side show a power-law dependence on $\omega$ given by [8]:

$$\left(\frac{\partial \ln G^*}{\partial p}\right)_{p=p_c} \sim \left(\frac{\partial \ln G^*}{\partial t_r}\right)_{t_r=t_c} \sim \omega^{-\kappa}, \tag{10}$$

where $\kappa$ is termed as the dynamic critical exponent while $t_r$ is the reaction time. This equation provides an important information about the gelation kinetics. The relationship given by Eq. (10) suggests as a system approaches the critical gel state the dynamic modulus evolves is such a fashion that the rate of change of moduli decreases with increasing frequency. The evolution of the properties is governed by the parameter $\kappa$ over all the timescales.

All the scaling exponents defined so far: $\alpha$, $n$, $s$, $z$, $\nu_S$, $\nu_G$ and $\kappa$ are positive constants specific to the system undergoing gelation. On correlating the critical exponents pertaining to the pre-gel and post-gel states independently, the following relationships given by [20]:

$$\nu_S = s/(1-n) \text{ and} \tag{11}$$

$$\nu_G = z/n, \tag{12}$$

are always observed to hold true. While approaching the critical gel state from the pre-gel side, the divergence of $\eta_0$ directly depends on the scaling of the rate of change of complex modulus given by Eq. (10). This leads to the interrelation between $s$, $n$ and $\kappa$ given by [6]:

$$s = (1-n)/\kappa. \tag{13}$$

Strictly speaking, the divergence of the relaxation time on either side of the critical gel state need not be correlated with each other. However, it has been postulated that the divergence of the dominant relaxation time is symmetric on both the sides of the critical gel point leading to [20]:

$$\nu_S = \nu_G. \tag{14}$$

Larsen *et al.* [21] studied peptide hydrogels and polyacrylamide gels using microrheology wherein they inferred the relaxation time from time-cure superposition that showed symmetric divergence of the longest relaxation. Such symmetry leads to a profound hyperscaling relationship given by [20]:

$$n = \frac{z}{z+s}, \tag{15}$$

which has been independently derived theoretically by various groups [22-24]. Such independent derivation of Eq. (15), as a result, implies symmetric divergence of the dominant relation time on either side of the critical gel point. On computing $n$ using Eq. (15) for randomly branched polyester [25], gelatin [26], chitosan [27], gellan gum [28] and $\kappa$ – carrageenan [29] systems at critical gel state, consistent results in $n$ with that obtained by Winter criteria were obtained.

Assuming that the postulate of symmetric divergence of dominant relaxation time on either side of the critical gel state given by Eq. (14) and its implication given by Eq. (15) holds true, the critical exponents $z$ and $n$ can be further related to the dynamic critical exponent $\kappa$ using the hyperscaling laws equations (13) and (15) as [8]:

$$z = n/\kappa. \tag{16}$$

Although equations (13) and (16) relates $\kappa$ to $n$, which itself varies in a limit $0 \leq n \leq 1$, for a large number of polymeric systems undergoing crosslinking reactions, $\kappa$ has been observed to be a constant with a value close to 0.2 [8,30,31]. In a seminal contribution, de Gennes [32] developed a relationship between $\alpha$ and $z$ given by:

$$z = 1 + (d-2)\alpha, \tag{17}$$

where $d$ is the dimension of space. The above mentioned scaling laws have been verified to some extent, though not entirely, for the chemically crosslinking polymeric gel systems, wherein the rheological measurements have been performed on the time-invariant state of a system obtained by terminating the crosslinking reaction at different stages in the vicinity of the critical gel point [4]. This methodology leads to a specific value of $\Delta p$ associated with various intermediate stages that remain constant during the rheological measurements. On the other hand, the scaling analysis has also been applied to the polymeric systems undergoing spontaneous chemical reactions [8,10,12,13]. Although there have been reports on validation of the hyperscaling law given by Eq. (15) for the polymeric gels obtained via chemical crosslinking [8,10,33,34], the postulated symmetric divergence of the dominant relaxation time, the experimental determination of $\kappa$ and its comparison with the prediction from scaling theories has not been studied in the literature for any kind of gel-forming system.

Apart from polymeric gels, the sol-gel transition has been widely observed in various kinds of physical gels that comprise colloidal as well as polymeric systems, which undergo sol–gel transition as a function of time [35-42] or as a function of temperature [29,43-49]. While the scaling laws serve well for chemical gels, the physical gels should be treated more meticulously. The evolution of viscoelastic properties near the sol–gel transition is studied by performing oscillatory rheological measurements on two different types of systems: a colloidal dispersion and a thermo-responsive polymer solution under isothermal and non-isothermal conditions. While undergoing sol–gel transition, both the systems pass through a critical point. An approach to the critical point is characterized in terms of divergence of zero shear viscosity and the subsequent appearance of the low frequency modulus. In the vicinity of the critical gel state, both the viscosity and the modulus show a power-law dependence on relative distance from the critical point. Interestingly, the longest relaxation time has been observed to diverge symmetrically on both the sides of the critical point and also shows a power-law dependence on relative distance from the critical point. The critical (power-law) exponents of the zero-shear viscosity and modulus are observed to be related to the exponents of the longest relaxation time by the hyper scaling laws. The dynamic critical exponent has also been calculated from the growth of the dynamic moduli. Remarkably, the critical relaxation exponent and dynamic critical exponent predicted from the scaling laws precisely agree with the experimental values from the isothermal as well as non-isothermal experiments. The associated critical exponents show remarkable internal consistency and universality for different kinds of systems undergoing the sol-gel transitioncarefully. It is still not clearly established as to which parameter will play a similar role to the extent of crosslinking $\Delta \tilde{p}$ in a chemical gelation process [50]. Recently, there has been some success in the application of the scaling laws for physical gels [27,51]. Rueb and Zukoski [52] observed the divergence of $\eta_0$ and $G_e$ in the vicinity of critical gelation temperature in the suspension of silica spheres with octadecyl chains. However, the estimation of $\eta_0$ was derived from a stress sweep test which is a destructive method. For the critical gels formed by gelatin, Guo *et al.* [26] observed the scaling law to be applicable only for a specific region of fast helix growth. Cho and Heuzey [27] verified the hyperscaling law given by Eq. (15) for a thermosensitive hydrogel of chitosan-$\beta$-glycerophosphate at different temperatures. Liu *et al.* [29] obtained a value of $\nu_S = 4.57$ and $\nu_G = 4.15$ using the experimentally determined value of $n$, $s$ and $z$ for an aqueous solution of $\kappa-$ carrageenan system. For a physical colloidal gel formed by silica particles, Negi and coworkers [53] were the first to experimentally report the value of $\kappa$. With the direct measurement of $n$ and $\kappa$, Negi and coworkers [53] predicted the value of $s$ and $z$ using Eq. (16) for the colloidal gel. A systematic estimation of $\kappa$ has also been carried out by Jatav and Joshi [37] for varying concentrations of synthetic hectorite clay and monovalent salt. Therefore, a careful investigation of the literature suggests that the scaling laws have been verified to some extent for the systems undergoing physical gelation. However, to the best of our knowledge, there is no comprehensive report wherein a single physical gel-forming system has been shown to obey all the scaling laws reported for the polymeric chemical gels. Furthermore, all the critical scaling exponents have not been measured experimentally for any gel forming system, including chemical and physical, undergoing a critical gel transition. Additionally, validity of the scaling laws under non-isothermal (temperature ramp) conditions has not been checked for any kind of gel forming system.

In this work, we study the linear viscoelastic behavior of two gel-forming systems namely a colloidal gel and a polymer gel, wherein network formation is due to the physical interactions. While the colloidal dispersion undergoes spontaneous time-dependent sol-gel transition, the polymer gel undergoes temperature-dependent sol-gel transition. The primary objective of this work is to critically analyze the applicability of the scaling laws, where the gelation is through the physical interactions, with time and temperature as the gelation variables. Secondly, we focus our attention on the effect of isothermal and non-isothermal conditions on the scaling law exponents. Lastly, we summarize the range of validity of the scaling laws and most importantly, we assess the interconnection amongst all the critical exponents at the critical gel point.

## II. Material and Experimental Procedure:

In this work, we study two gel-forming systems, namely, an aqueous dispersion of synthetic hectorite clay and aqueous solution of Poly (vinyl) alcohol (PVOH). The first system is an aqueous dispersion of LAPONITE® XLG (a registered trademark of BYK Additives), which is synthetic hectorite clay mineral. The individual clay particle of LAPONITE® XLG is disk-like with an average diameter of around 25 nm and a thickness of 1 nm [54]. Hectorite mineral belongs to the family of 2:1 layered silicate, where one octahedral sheet of magnesia is sandwiched by two tetrahedral sheets of silica [5]. The unit cell formula of the synthetic hectorite clay is given by: $\text{Na}_{+0.7}\left[\left(\text{Si}_8\text{Mg}_{5.5}\text{Li}_{0.3}\right)\text{O}_{20}\left(\text{OH}\right)_4\right]_{-0.7}$. In an aqueous media, hectorite particle acquires the dissimilar charges with faces and edges possessing respectively the negative and the positive charge. As a result, the particles share attractive and repulsive interactions in an aqueous medium eventually forming an edge to face attractive bond leading to spontaneous sol-gel transition [37,55]. In this work, we study 3 weight % LAPONITE® XLG, 2 mM NaCl dispersion by mixing oven-dried (120 °C for 4 h) clay in ultrapure water from Millipore purification system (resistivity 18.2 MΩ.cm) having a predetermined quantity of salt. The stirring is carried using IKA Ultra-Turrax Drive for a duration of 30 minutes, which leads to a clear homogeneous dispersion. The freshly prepared samples are used for independent rheological measurements that are carried out immediately after the stirring is stopped.

Poly (vinyl alcohol) (PVOH) with molecular weight 89000 – 98000 g/mol is procured from Sigma-Aldrich. It is a semi-crystalline polymer which forms physically crosslinked poly (vinyl alcohol) hydrogels with temperature [56]. We prepare 10 weight % PVOH solution in deionized Millipore water. A predetermined quantity of PVOH is dissolved in ultrapure Millipore water at 75 °C under continuous stirring at 300 rpm for 8 hours using an IKA C-MAG-HS hot plate magnetic stirrer. In order to prevent any evaporation losses, the stirring is conducted in a closed container. The freshly prepared clear suspension of PVOH is then allowed to cool slowly and stored for 4 hours under ambient conditions. The colloidal dispersion of hectorite clay undergoes a spontaneous sol-gel transition as a function of time, while aqueous PVOH solution undergoes sol-gel transition upon decreasing the temperature.

Rheological experiments on colloidal dispersion and polymer solution were carried out using a Dynamic Hybrid Rheometer 3 (TA Instruments) with smooth concentric cylinder geometry (cup diameter 30 mm and gap 1 mm). We employ time-resolved rheometry to investigate the sol-gel transition in the colloidal dispersion and polymer solution. Subsequent to loading of a sample into the shear cell, the colloidal dispersion is subjected to a cyclic frequency sweep experiment, conducted at an oscillatory stress of 0.1 Pa over an angular frequency range of 0.5 – 25 rad/s at constant but different temperatures ranging between 10 °C to 50 °C. We also carry out cyclic frequency sweep experiments on colloidal dispersion under the application of continuous temperature ramp rates from 0.05 to 0.4 °C/min. In the case of PVOH solution, after a sample is loaded in the rheometer geometry, the temperature is decreased from 30 °C to 0 °C at the specified cooling rates ranging between –0.05 to –0.4 °C/min. During the process of cooling, the sample is continuously subjected to cyclic frequency sweep in the range of 0.5 – 25 rad/s with a stress magnitude of 1 Pa. In order to have accurate and acceptable results of oscillatory experiments, the frequency range is chosen in such a way that the mutation numbers $\left(N'_{mu} = \left(2\pi/\omega G'\right)\left(\partial G'/\partial t\right)\right)$ and $\left(N''_{mu} = \left(2\pi/\omega G''\right)\left(\partial G''/\partial t\right)\right)$ remain within the stated limits: $N'_{mu} < 0.1$ and $N''_{mu} < N'_{mu}$ [57]. Furthermore, we do not explore very low frequencies as it would require long experimental times during which the sample structure might change. All the experiments have been repeated atleast twice to ensure the repeatability of the results. In all the experiment, the temperature is controlled using the Peltier plate assembly and a thin layer of silicon oil is applied on the free surface of the sample to prevent any contamination and evaporation losses.

## III. Results and Discussions:

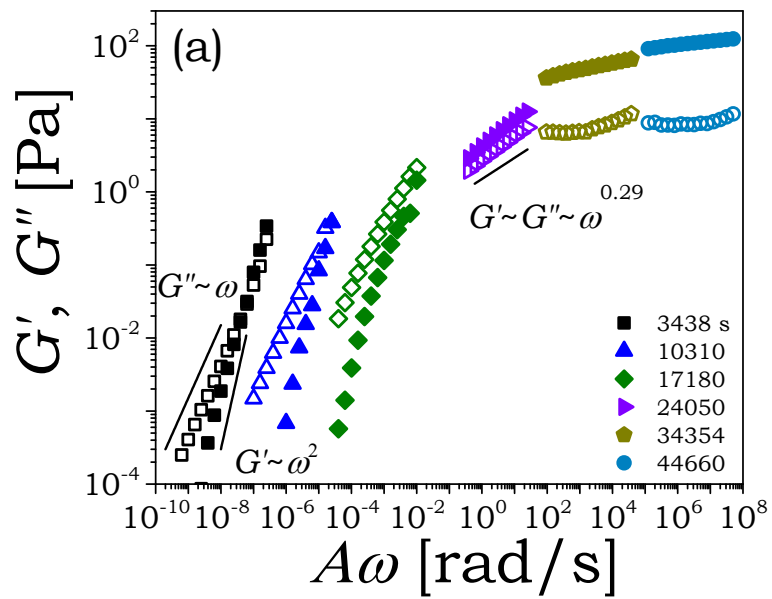

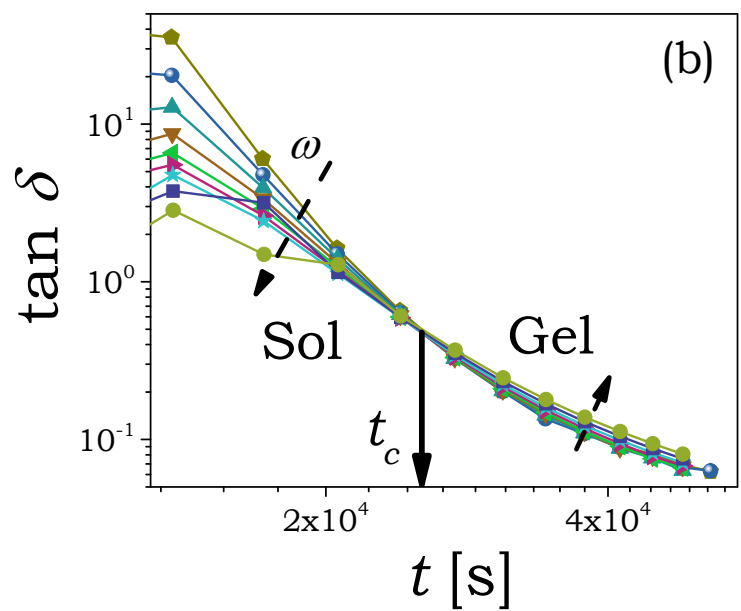

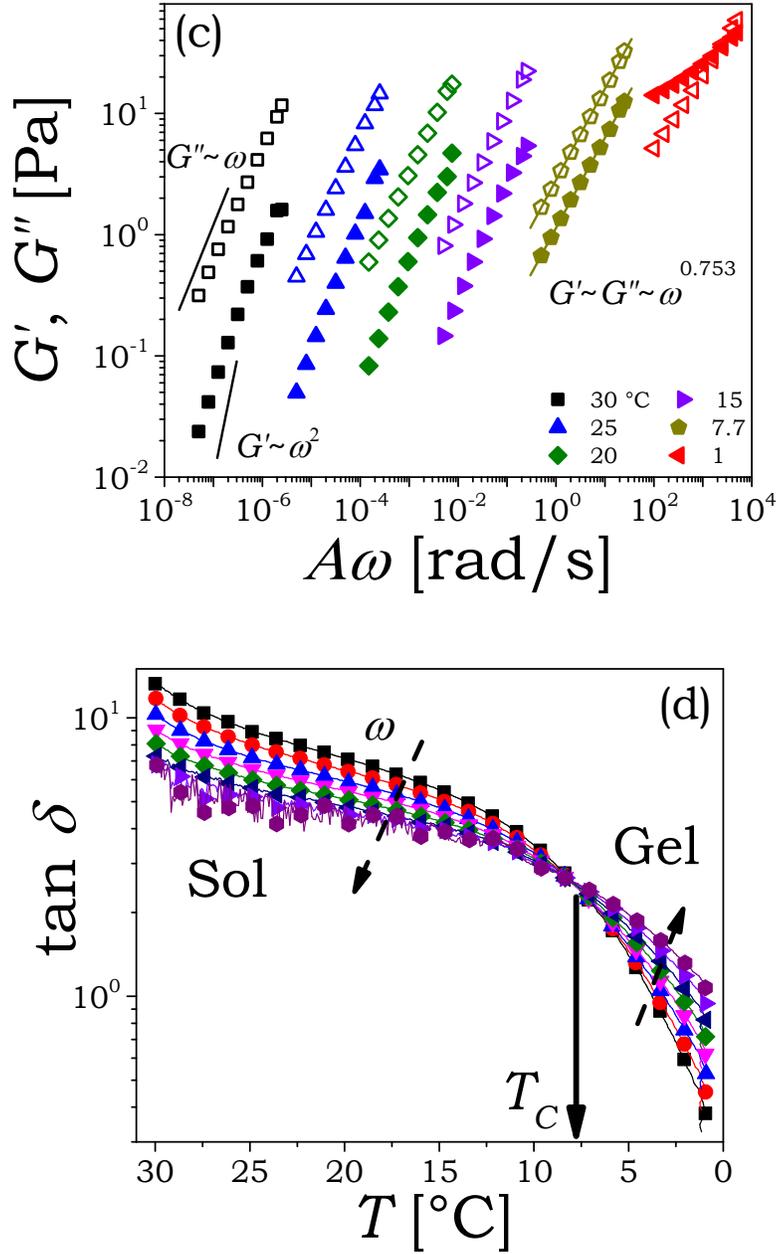

**FIG. 1.** The dynamic moduli $G'$ (closed symbols) and $G''$ (open symbols) are plotted as a function of frequency $(\omega)$ at different stages of gelation for (a) the colloidal dispersion of hectorite clay at 30°C and for (c) PVOH solution undergoing cooling at a rate of 0.05°C /min. The horizontal axis in (a) and (c) has been shifted by a factor $A$ for clarity. The variation in $\tan\delta$ is plotted as a function of time for (b) the colloidal dispersion and (d)

the polymeric solution during the gelation process. The dashed arrow indicates the direction of increasing value of angular frequency. The solid arrow marks the point of the critical gel transition. The horizontal axis in figure (d) has been drawn in decreasing order of magnitude of temperature to emphasis the cooling process.

We perform oscillatory shear experiments on the colloidal dispersion of synthetic hectorite clay and the PVOH solution over the two decades of angular frequency $(\omega)$ as these systems undergo the microstructural evolution respectively as a function of time and temperature. The time evolution of frequency dependence of $G'$ and $G''$ in colloidal dispersion is plotted in Fig. 1(a) while the variation of the same is plotted at different temperatures for the PVOH solution in Fig. 1(c). At the beginning of the experiment, both the systems are liquid-like as described by the frequency dependence of $G' \sim \omega^2$ and $G'' \sim \omega$, with $G'' >> G'$, which signifies the systems to be in a sol state. With an increase in time for the clay dispersion and decrease in temperature for PVOH solution, $G'$ evolves faster than $G''$ until a specific point when both the moduli display an identical dependence on $\omega$, $G' \sim G'' \sim \omega^n$, as given by Eq. (1). Such identical dependence of moduli serves as the rheological signature of a critical gel state. With further increase in time for the clay dispersion and decrease in temperature for PVOH solution, the dependence of $G'$ on frequency continues to weaken, eventually showing a plateau in the low-frequency limit. In Fig. 1(b) (colloidal dispersion) and 1(d) (PVOH solution), we plot the variation of the $\tan \delta$ respectively as a function of time and temperature. The dashed arrow in Fig. 1(b) and (d) signals the direction of increase in $\omega$. As expected, in the sol state, $\tan \delta$ decreases with an increase in $\omega$. However, at a certain time for clay dispersion and temperature for PVOH solution, all the iso-frequency $\tan \delta$ curves cross at an identical point. This point is termed as the critical gel point. Beyond the critical gel state, $\tan \delta$ increases with an increase in $\omega$ that is suggestive of a post-critical gel state. Figure 1, therefore suggests that both the systems show all the characteristic features of a sol-gel transition with regard to the respective controlling variables. The corresponding time (for clay dispersion) and temperature (for PVOH solution) associated with the critical gel state is shown as a solid arrow in Fig. 1(b) and (d), which is represented as $t_c$ and $T_c$ respectively. Furthermore, the knowledge of $\tan\delta$ at the point of intersection leads to the estimation of the critical relaxation exponent as $n = 2\delta/\pi$ [40]. The obtained value of $n$ is identical to the power-law exponent of the moduli dependence on $\omega$ as described by Eq. (1) [58]. Therefore, Fig. 1 clearly establishes the presence of a critical gel state, where the colloidal dispersion and PVOH solution form the weakest space spanning percolated network corroborating with the Winter criteria [14,15].

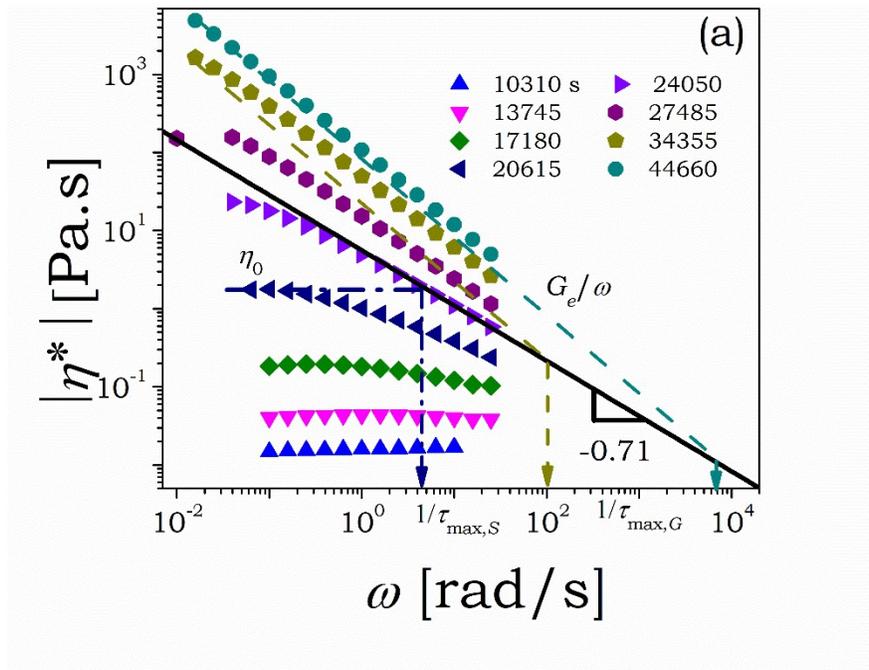

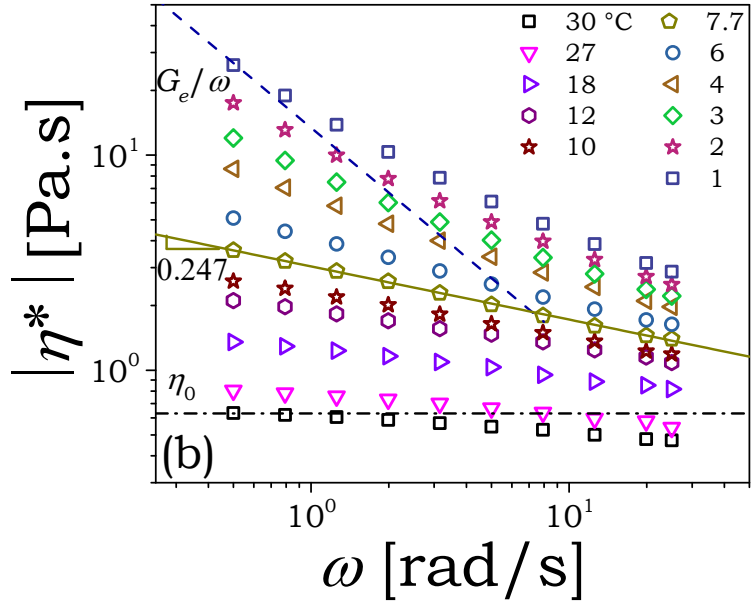

**FIG. 2.** Complex viscosity $|\eta^*|$ is plotted as a function of $\omega$ at different extents of gelation in (a) the colloidal dispersion of hectorite clay and (b) the PVOH solution. The solid line denotes the power-law decay of $|\eta^*|$ with an exponent of -0.71 and –0.247 respectively for the clay dispersion and the polymer gel at the critical state as represented by Eq. (18). The low $\omega$ asymptotic behavior is shown by the dash-dotted line (given by the expression, $|\eta^*| = \eta_0$) in the pre–gel region with zero shear viscosity $\eta_0$ as a plateau value. The dashed lines indicate the low $\omega$ asymptotes (given by the expression: $|\eta^*| = G_e/\omega$) for the post – gel states. The reciprocal of frequency associated with the intersection of low $\omega$ asymptote and the critical gel state (indicated by the arrows) represents the characteristic relaxation time in the pre-gel and post-gel states as discussed in the text.

In addition to detecting the critical gel state, the results of the dynamic mechanical measurements shown in Fig. 1 enable estimation of other rheological properties such as viscosity, equilibrium elastic modulus and relaxation time during the sol-gel transition, wherein the critical state serves as the reference state. In order to get further insight into the evolution of viscoelastic behavior, we plot complex viscosity $|\eta^*|$ as a function of $\omega$ at different times elapsed since sample preparation in Fig. 2(a) for the clay dispersion. Initially, the clay dispersion is liquid-like, which results in a low value of $|\eta^*|$ that is independent of $\omega$. In order to avoid misinterpretation of the oscillatory shear data for very low viscosity samples, we compute the complex dimensionless Stokes number $(i\rho\omega l^2)/\eta^*$ where $\rho$ is the density of the fluid and $l$ is the gap thickness [59]. According to Bird *et al.* [60], the impact of fluid inertia on the flow can be neglected if the condition $\left|(i\rho\omega l^2)/\eta^*\right|^{1/2} < 1$ holds true. For the very early times of the gelation process $(t < 7000\text{s})$, the measured value of $|\eta^*|$ is of the order of $10^{-3}$ Pa.s, which results in a higher value of the Stokes number $\left(\left|(i\rho\omega l^2)/\eta^*\right|^{1/2} > 1\right)$ for $\omega > 1$ rad/s. Therefore, we show the data only for $t > 7000$ s. As the value of $|\eta^*|$ increases as the gelation progresses, the data becomes more reliable at lower $\omega$. In Fig. 2(b), we plot the evolution of $|\eta^*|$ for PVOH dispersion as a function of $\omega$ at different temperatures during cooling. It is important to note that the cyclic frequency sweep experiment is carried out on a time-evolving system for the colloidal dispersion and temperature-evolving system for the PVOH solution. Therefore, the time associated with each point on the plot pertaining to the colloidal dispersion and temperature corresponding to the PVOH solution is different. However, for convenience, in the legends, we denote all the lower frequency data by the same time/temperature, which is associated with the highest frequency of $\omega = 25$ rad/s. In all the experiments, we assume that the evolution of the sample over a single frequency sweep is not significant.

With an increase in time, the clay particles interact via edge to face bonds and form the clusters [5]. These clusters grow with time leading to an increase in $|\eta^*|$ as shown in Fig. 2(a). It can be seen that in the pre-gel state, $|\eta^*|$ exhibits a plateau in a limit of small $\omega$ and decreases with $\omega$ beyond a certain $\omega$. The onset of $\omega$ at which $|\eta^*|$ starts decreasing, shifts to lower value at a greater extent of gelation. For the aqueous PVOH solution, wherein gelation occurs due to decrease in temperature, the polymer chain segments form hydrogen bonds thereby forming a junction point. The increase in the number of segments participating in the hydrogen bonding network with a decrease in temperature results in enhanced value of $|\eta^*|$. Since the sol state obeys the Cox-Merz rule [61], the plateau value of $|\eta^*|$ in the limit of low $\omega$ leads to the estimation of zero shear viscosity as per $\eta_0 = \lim_{\omega \to 0} |\eta^*|$, as indicated by the dashed-dotted line in Fig. 2(a) and (b).

With further increase in the gelation time, the growing clusters formed by the anisotropically charged clay particles are just able to form a space spanning percolated network represented as the critical gel state. Equivalently, as more polymer segments engage in the hydrogen bonding, a point comes when these junctions, connected by polymer segments lead to the weakest percolated network. At the critical gel point, $|\eta^*|$ exhibits a power-law dependence on $\omega$ given by [20]:

$$\eta^*(\omega, \varepsilon) = aS(i\omega)^{n-1} \text{ such that } |\eta^*(\omega, \varepsilon)| = aS\omega^{n-1}, \tag{18}$$

where $a = \pi / (\Gamma(n) \sin(n\pi))$. The prediction of Eq. (18) for $n$ and $S$ obtained from Fig. 1 is plotted for the colloidal gel of clay and that of PVOH solution in Fig. 2(a) and (b) respectively. It can be seen that Eq. (18) fits the experimental data in Fig. 2(a) and (b) very well. With an increase in time for the clay dispersion and decrease in temperature for aqueous PVOH solution, the respective networks get more consolidated due to the strengthening of the junction points and/or enhancement of the junction density. In the post-gel state, as shown in Fig. 2(a) and (b), $|\eta^*|$ grows further with gel consolidation, and in a limit of small $\omega$ diverges with an inverse $\omega$ dependence that is suggestive of a solid-like response. The dependence of $|\eta^*|$ in the low $\omega$ region facilitates estimation of the equilibrium modulus ($G_e$) given by: $|\eta^*| = \lim_{\omega \to 0} G_e(t, \varepsilon) / \omega$. The low $\omega$ asymptotes for a pre-gel (20615 s) and post-gel (34354 s and 44658 s) are constructed as dash-dotted and dashed lines respectively in Fig. 2(a). These asymptotes intersect with the power-law of the critical state such that the inverse of $\omega$ associated with the intersection point defines the longest relaxation time $\tau_{\max,S}$ in the pre-gel state and $\tau_{\max,G}$ in the post-gel state of the gel-forming system. In a limit of high $\omega$, the $|\eta^*|$ dependence on $\omega$ approaches the power law with the same exponent as that of the critical gel state.

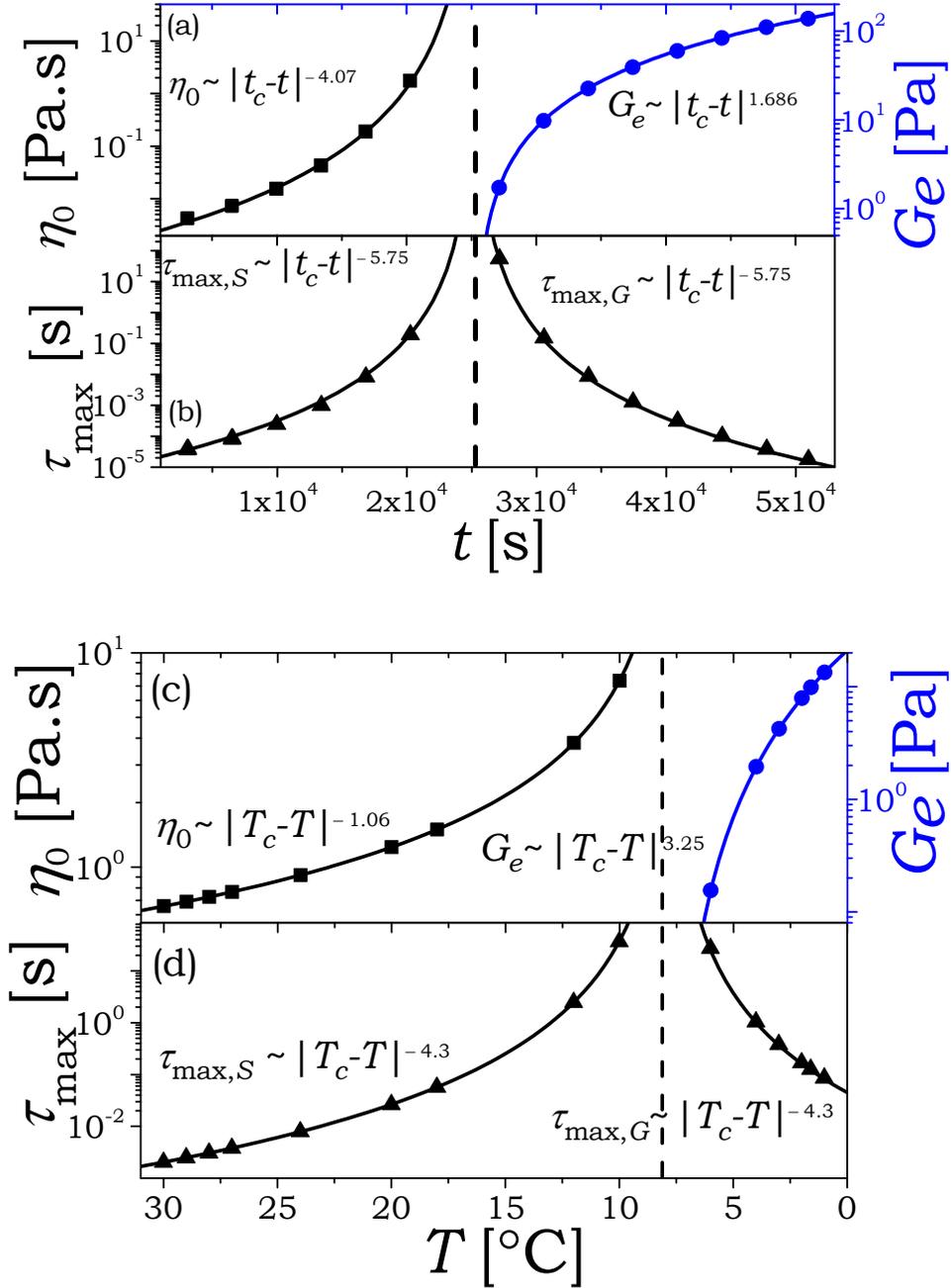

**FIG. 3.** Evolution of $\eta_0$, $G_e$ and $\tau_{\max}$ is plotted as a function of time for colloidal dispersion of hectorite clay (a and b) and temperature for PVOH solution (c and d). The critical gel point is represented by the dashed line. The solid lines denote the scaling laws for $\eta_0$, $G_e$ and $\tau_{\max}$ mentioned and discussed in the text.

In Fig. 3, we plot the evolution of $\eta_0$ and $\tau_{\max,S}$ in the pre-gel while $G_e$ and $\tau_{\max,G}$ in the post-gel state as a function of time for the colloidal gel (Fig. 3(a) and (b)) and as a function of temperature for the PVOH gel (Fig. 3(c) and (d)). The critical gelation point is shown as a dashed vertical line in Fig. 3. In the pre-gel state, the clusters grow in size resulting in enhancement of $\eta_0$ and $\tau_{\max,S}$. At the critical gel state, the percolated network spans the space as the cluster size approaches infinity causing $\eta_0$ and $\tau_{\max,S}$ to diverge. Beyond the critical gel point, the solid-like behavior of the network is characterised by the growing value of equilibrium modulus (in a limit of zero frequency) $G_e$. The magnitude of $G_e$, which is zero at the critical point, increases in the post-gel regime as the percolated network consolidates. Similarly, the divergence of $\eta_0$ and $\tau_{\max,S}$ is observed for the PVOH gel in the vicinity of the critical gel state and $G_e$ starts increasing as the temperature is lowered below the critical temperature. For both the systems, in the post-gel state, $\tau_{\max,G}$ belongs to the relaxable components (free or unattached clusters, or the PVOH chains that are not part of any junction). During the consolidation of the gel, these relaxable components progressively participate in the network such that the large free clusters get attached to the network faster than the smaller ones. Consequently, beyond the critical point, $\tau_{\max,G}$ decreases with an increase in time for the colloidal gel and decrease in temperature for the PVOH gel.

As discussed in the introduction, the zero shear viscosity $\left(\eta_0\right)$ and the equilibrium modulus $\left(G_e\right)$ are observed to scale with $\Delta \tilde{p}$ for a polymeric gel undergoing chemical crosslinking reaction [20,62]. Since the systems explored in this work undergo physical gelation, it is not possible to obtain $\Delta \tilde{p}$ as the gelation progresses. As the colloidal dispersion transforms from a sol to a gel state with change in time, in accordance with that proposed by Winter and coworkers [8], we define $\varepsilon = |t_c - t|/t_c$ as the dimensionless relative distance from the gelation threshold in terms of time. On similar grounds, for PVOH solution, where the sol-gel transition is induced by a decrease in temperature, we propose $\varepsilon$ can be defined in terms of temperature as $\varepsilon = |T_c - T|/T_c$. By analogy with the chemical gelation, we fit the scaling laws: $\eta_0 \sim \varepsilon^{-\bar{s}}$, $G_e \sim \varepsilon^{\bar{z}}$, $\tau_{max,S} \sim \varepsilon^{-\bar{\nu}_S}$ and $\tau_{max,G} \sim \varepsilon^{-\bar{\nu}_G}$ to the experimental data for the colloidal gel and the PVOH gel. The lines shown in Fig. 3 describe the fits to the above mentioned scaling laws that show an excellent agreement with the experimental data. This leads to the estimation of the different scaling exponents for the studied physical gels. The values of the critical exponent have been tabulated in Table 1.

With the knowledge of $\bar{s}$, $\bar{z}$, $\bar{\nu}_S$ and $\bar{\nu}_G$, we compute the critical relaxation exponent from the pre-gel and post-gel side ($n_S$ and $n_G$) using the equations equivalent to Eq. (11) and (12) represented by: $n_S = 1 - (\bar{s}/\bar{\nu}_S)$ and $n_G = \bar{z}/\bar{\nu}_G$. Remarkably, for the colloidal gel, we get $n_S = n_G = 0.29$, which is identical to the value of $n$ obtained from the Winter Chambon criteria from the analysis of Fig. 1. Likewise for PVOH gel, the value of $n$ matches very well with the exponents ($n_S = n_G = 0.753$) computed using the scaling laws. This result, therefore, indicates that the approach to the critical gel point from the sol side is identical to that of from the post-gel side. An equivalence of $n = n_S = n_G$ for both the systems suggests that $\Delta \tilde{p} \sim \varepsilon$. As discussed in the introduction, such equivalence has been reported for the chemically crosslinking polymeric systems undergoing spontaneous gelation with time [8,10,12]. Furthermore, $\Delta \tilde{p} \sim \varepsilon$ has been reported to be valid, when the rate of formation of the crosslinks is first order with respect to the participating reactants [9]. This suggests that the process of gelation in the explored systems is first order with respect to the corresponding gelation variable $(X)$: increase in $t$ for colloidal dispersion and decrease in $T$ for PVOH solution. With an objective to understand the implications of this profound statement, we consider $[M]$ as the concentration of the precursor units that have the potential to participate in crosslink and $[N]$ as the concentration of the junction nodes where the individual precursor units will participate to form a percolated network. On correlating the process of physical gelation with chemically crosslinking reaction, the rate of change in $[M]$ with respect to the corresponding gelation variable $X$ in the neighborhood of critical state can be mathematically expressed as: $\frac{d[M]}{dX} = -k[M][N]$. Assuming that in the vicinity of the critical gel state, $[N]$ remains constant, the above relation reduces to: $\frac{d[M]}{dX} \sim k[M]$, which clearly suggests that with as the critical gel state is approached, the number of precursor units gets consumed exponentially with the extent of gelation. The equivalence $\Delta \tilde{p} \sim \varepsilon$ is further corroborated by the symmetric divergence of the dominant relaxation time in the vicinity of the critical gel point for both the systems, which leads to $\bar{\nu}_S = \bar{\nu}_G$ as shown in Fig. 3(b) and (d). This is an important result as it experimentally validates the symmetric evolution of the relaxation time on either side of the critical gel state and the hyperscaling laws for the two different types of physical gels with distinct gelation variables. Many experimental studies have also demonstrated the validation of hyperscaling law given by Eq. (15) for different gelling systems [28,29,53] as tabulated in Table. 1. Furthermore, with the knowledge of $\bar{z}$, the exponent associated with the divergence of the largest cluster size $(\alpha)$ can be computed using the relationship proposed by de Gennes (Eq.(17)) for three-dimensional space. This suggests the characteristic size of the largest cluster grows as $\xi \sim \varepsilon^{-0.68}$ for the colloidal dispersion and $\xi \sim \varepsilon^{-2.2}$ for the PVOH solution. It is evident that the cluster size grows very sharply near the critical gel point in case of PVOH solution than in colloidal dispersion. We believe that the origin of such behavior is determined by the nature of gelation in these systems. In case of dispersion of hectorite, the network formation appears to be more gradual at all times. The hydrogen bonding network associated with the PVOH chains, on the other hand, seems to be growing more rapidly with a decrease in the temperature in the vicinity of the critical point.

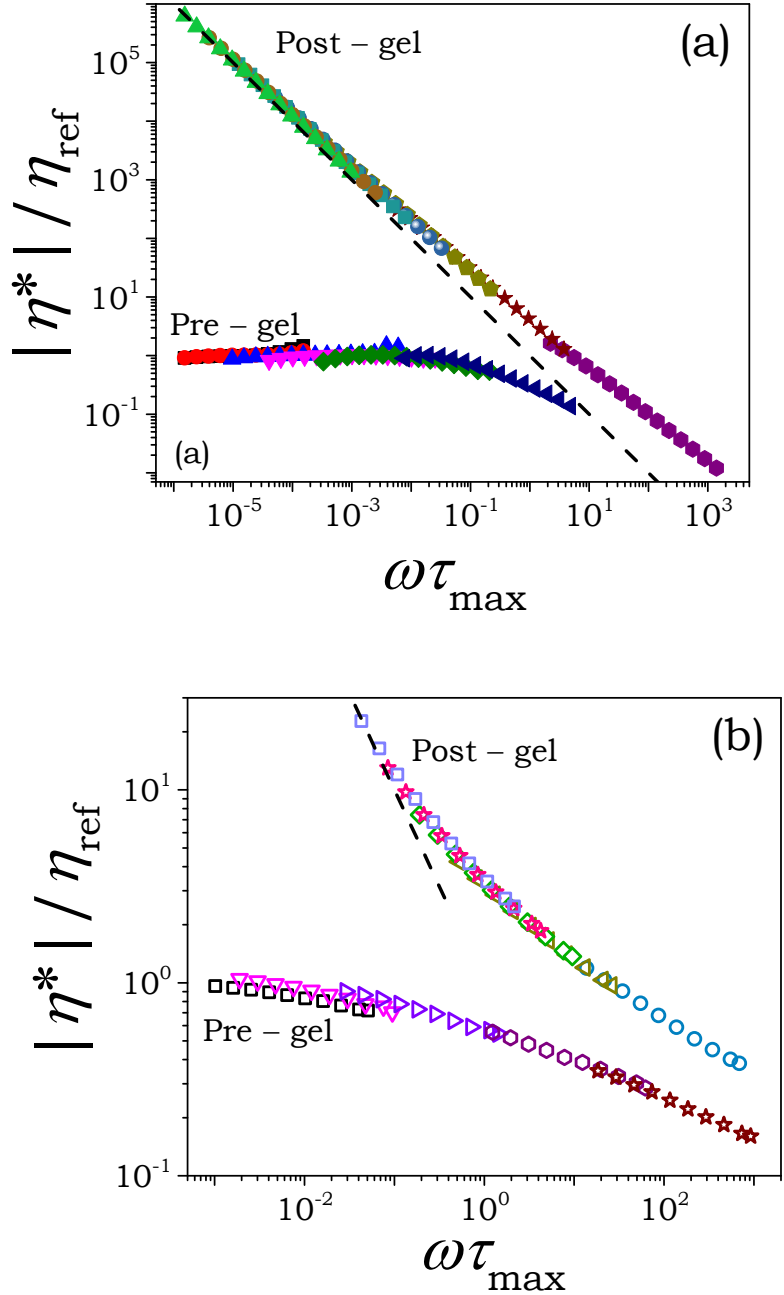

**FIG. 4.** Normalized $|\eta^*|$ is plotted as a function of $\omega\tau_{max}$ for dispersion of hectorite clay (a) and PVOH solution (b). The ordinate in both the plots is normalized by reference value. The value of $\eta_{ref}$ in the pre – gel region is $\eta_0$ while in the post – gel region it is taken as $G_e \tau_{max,G}$. Owing to the divergence of relaxation time at the critical gel state, data associated with the same cannot be shown on this plot. The dashed line denotes a slope of $-1$.

With an estimation of $\eta_0$ and $G_e$, for the data shown in Fig. 3, $|\eta^*|$ scaled using a reference viscosity $(\eta_{ref})$ is plotted as a function of $\omega \tau_{max}$ in Fig. 4(a) and (b) respectively for the colloidal dispersion and PVOH solution. The value of $\eta_{ref}$ is taken as $\eta_0$ and $G_e \tau_{max,G}$ while $\tau_{max}$ is taken as $\tau_{max,S}$ and $\tau_{max,G}$ in the pre-gel and the post-gel states respectively. It can be seen that viscosity data for both the gel-forming systems collapse very well into two domains for the pre-gel and the post-gel states giving rise to two different shapes of the curves. In the pre-gel state, all the shifted viscosity curves display a constant value of normalized viscosity followed by a frequency-dependent decrease in the same. On the other hand, the collapsed post-gel complex viscosity data decay with a slope of $-1$ in a limit of low frequency as indicated by the dashed line in Fig. 4(a) and (b). The high-frequency behavior of the pre-gel and the post-gel states seem to be approaching the power-law behavior of the critical state. Figure 4, therefore, leads to superposition in the entire domain of sol-gel transition for both the systems. Superposition of the complex viscosity independently in the pre-gel and the post-gel state when the frequency is normalized by the dominant relaxation time implies that while the relaxation time diverges on either side of the critical gel state, the shape of the relaxation time distribution remains preserved independently in both the pre as well as the post gel state. Interestingly preservation of the shape of relaxation time distribution on either side of the critical gel point has also been reported by Kaushal and Joshi [63] for a PDMS system undergoing sol-gel transition by using effective time domain approach. While the shape of relaxation time distribution independently remains preserved in the pre and the post critical gel state domains, the shape at the critical gel state always remains unique given by Eq. (4). Additionally, the conservation of the shape of relaxation time spectrum is corroborated using microrheology by Furst and coworkers [21,64,65] for peptide hydrogel, polyacrylamide gels and PEG–heparin hydrogels wherein they obtained time-cure superposition by arbitrarily shifting the individual mean-squared displacement curves in pre-gel and post-gel regimes. Interestingly, the inverse of the shift factor used for time reflects the longest relaxation time which has been observed to diverge symmetrically.

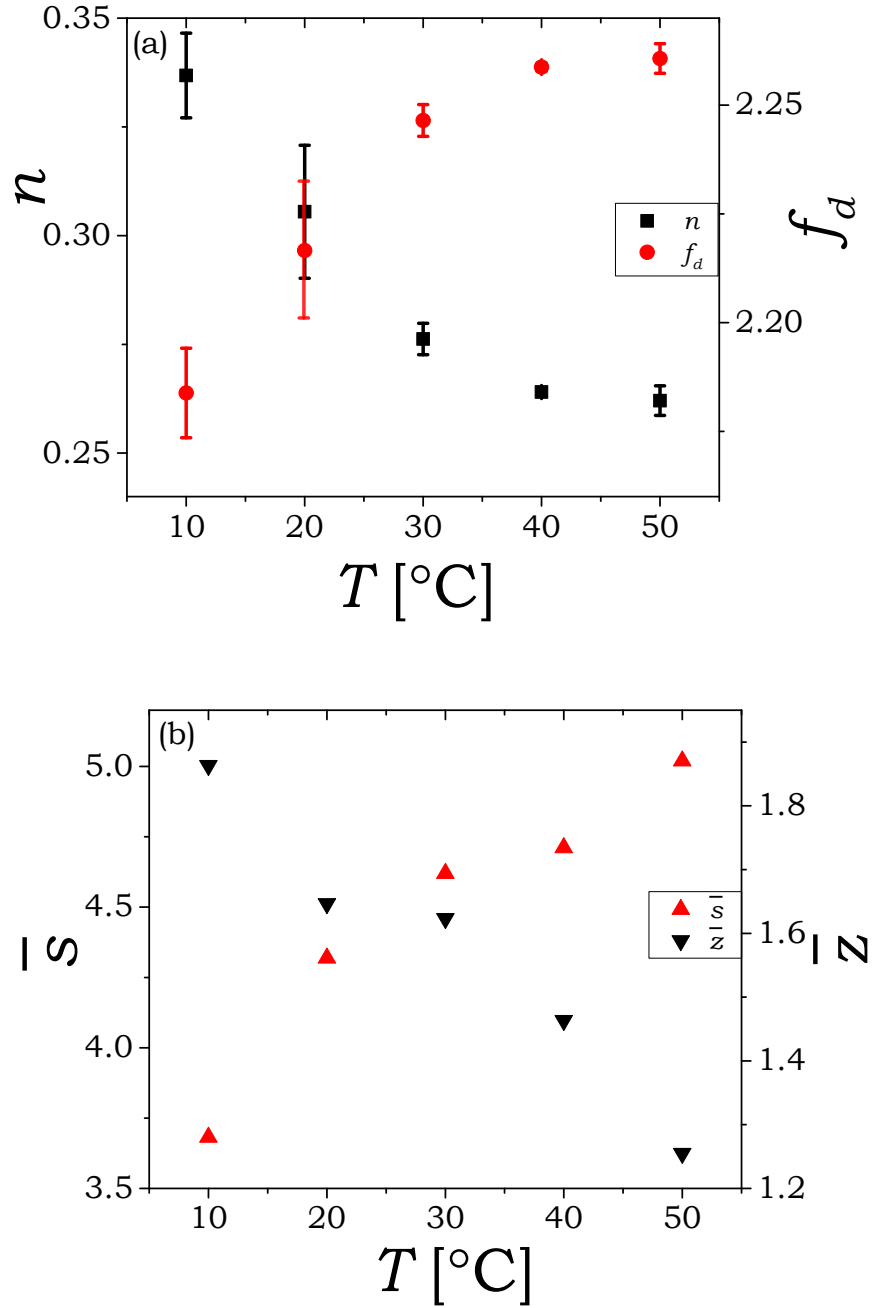

**FIG. 5.** Variation of (a) relaxation exponent $(n)$ and fractal dimension $(f_d)$ and the scaling exponents (b) $\bar{s}$ and $\bar{z}$ are plotted as a function of temperature for the colloidal dispersion of hectorite clay.

We also investigate the sol-gel transition in the colloidal clay dispersion at different temperatures in the range of 10°C – 50 °C. The corresponding evolution of linear viscoelastic parameters leads to the estimation of the exponents associated with the critical gel state. The process of gelation in the aqueous clay dispersion is known to get accelerated with an increase in the temperature causing the time at which the critical gel state is observed to decrease [37]. In Fig. 5(a), we plot $n$ associated with the individual isothermal experiments as a function of temperature. It can be seen that the value of $n$ decreases with accelerated gelation kinetics. On the contrary, for the polymeric gel, Winter and co-workers [14,66] showed that the value of $n$ remains the same over a wide range of temperatures that allow time-temperature superposition. For the colloidal gel of hectorite clay studied in the present work, since $n$ varies with temperature, the time-temperature superposition cannot be applied. Assuming the fractal structure of the critical colloidal gel is analogous to an arbitrarily branched polymer melt system with a complete screening of the excluded volume effects, the critical exponent $n$ leads to the fractal dimension $(f_d)$ given by $f_d = 5(2n_c - 3)/2(n_c - 3)$ [17]. The computed value of $f_d$ is plotted as a function of temperature in Fig. 5(a). It can be seen that with an increase in the temperature, the decrease in $n$ corresponds to enhancement in $f_d$. This suggests that with the increase in the temperature, the enhanced thermal energy in the system causes the particles to make more number of attempts before forming the interparticle bonds leading to more dense cluster through reaction-limited cluster aggregation (RLCA) [58,67,68].

We also obtain the critical scaling exponents associated with the evolution of zero shear viscosity and equilibrium modulus respectively $\bar{s}$ and $\bar{z}$ at different temperatures, and are plotted in Fig. 5(b). It can be seen that $\bar{s}$ increases while $\bar{z}$ decreases with increase in the temperature. Increase in $\bar{s}$ suggests that $\eta_0$ diverges at a faster rate with an increase in temperature as it approaches the critical gel point. As a greater number of clusters gets formed due to enhanced gelation kinetics, the value of $\eta_0$ achieves a higher value at elevated temperatures. This leads to a higher growth rate of $\eta_0$, governed by $\bar{s}$, with an increase in temperature. On the other hand, a decrease in $\bar{z}$ suggests the growth rate of $G_e$ in the post-gel state decreases with increase in the temperature. We shall discuss the implications of both these behaviors below.

The introduction section also discusses an additional critical exponent $\kappa$ that describes frequency dependence of evolution of complex moduli with respect to the degree of crosslinking $p$ given by Eq. (10). For the other scaling laws, since we find $\varepsilon$ for physical gelation to be equivalent to $\Delta \tilde{p}$ for chemical gelation, Eq. (10) can be written as:

$$\left(\frac{\partial \ln G^*}{\partial p}\right)_{p=p_c} \sim \left(\frac{\partial \ln G^*}{\partial X}\right)_{X=X_c} \sim \omega^{-\kappa}, \tag{19}$$

where $X$ is the respective gelation variable, time or temperature correspondingly for the colloidal gel or PVOH gel. Since the derivative is computed in close vicinity of the critical gel state, the independent variable $p$ can be replaced by $t$ or $T$ such that $\Delta p \sim \Delta t$ for a spontaneously evolving system while $\Delta p \sim \Delta T$ for temperature-dependent gel-forming system [8]. Furthermore, on representing $G^*$ in terms of $G'$ and $G''$, Eq. (19) can be written as [8,53]:

$$\left(\frac{\partial \ln G'}{\partial X}\right)_{X=X_c} = C \left(\frac{\partial \ln G''}{\partial X}\right)_{X=X_c} \sim \omega^{-\kappa}, \tag{20}$$

where $C$ is the constant of proportionality. Using the oscillatory shear data shown in Fig. 1(a) and (c), the values of $\kappa$ and $C$ can be estimated for both the gel-forming systems.

In order to compare the equivalence of Eq. (19) and (20), we compute the value of $\kappa$ using the first derivative of $G^*$, $G'$ and $G''$ with respect to $X$ independently. For the data shown in Fig. 1(a) and (c), the computed value of $\kappa$ using the first derivative of $G^*$, $G'$ and $G''$ results in similar value. Therefore, throughout this work, we report the value of $\kappa$ computed using $G^*$. Furthermore, using Eq. (20) we obtain $C = 2.06$ for the clay dispersion and $C = 2.32$ for the PVOH solution. Interestingly, the computed value of $\kappa$ and $C$ are similar to the critical constants reported for various polymeric gels [8,29,30,69], thus underlining a similarity between the diverse physical gelation processes studied in this work and the chemical gelation.

Equations (10) and (19) can also be represented in terms of relaxation modulus $\big(G(t)\big)$, wherein reciprocal of frequency is replaced by time and is given by [8]:

$$\left(\frac{\partial \ln G(t)}{\partial p}\right)_{p=p_c} \sim t^{\kappa}. \tag{21}$$

Scanlan and Winter [8] proposed that in the pre-gel state $G(t)$ follows a dependence given by:

$$G(t,p) = St^{-n} \exp\left(\zeta \Delta p t^{\kappa}\right), \tag{22}$$

where $\zeta$ is a constant specific to a material. This suggests that the shape of relaxation time spectra remains unaffected by the extent of gelation in the pregel state, which is also apparent from the superposition of the complex viscosity shown in Fig. 4. The zero shear viscosity in the pre-gel state can easily be obtained by integrating Eq. (22) as:

$$\eta_0(p) = \int_0^{\infty} G(t,p)dt = S\int_0^{\infty} t^{-n} \exp\left(\zeta \Delta p t^{\kappa}\right)dt = \frac{S}{\kappa}\Gamma\left(\frac{1-n}{\kappa}\right)\left(\zeta \Delta p\right)^{(n-1)/\kappa}, \tag{23}$$

which for the present systems is analogous to $\eta_0 \sim \varepsilon^{(n-1)/\kappa}$. This relationship suggests $\kappa = (1-n)/\overline{s}$. Even in the post-gel state, superposition of the complex viscosity shown in Fig. 4 suggests that the shape of relaxation time spectra remains preserved, although no analytical relation is available for the same that is equivalent to Eq. (22) for the pre-gel states. However, since the evolution of relaxation time is symmetric on both the sides of critical gel state leading to the relationship among $\bar{s}$, $\bar{z}$ and $n$, an equivalent relation can also be obtained for the post-gel as: $\kappa = n/\bar{z}$. Very interestingly, on comparing these relations with Eq. (11) and (12), we get $\nu_S = \nu_G = \kappa^{-1}$. With a knowledge of experimentally obtained $\kappa$ from the derivative of complex moduli given by Eq. (19), we assess $\kappa = (1-n)/\bar{s}$ by comparing it with the value of $\kappa$ derived from $\bar{s}$, $\bar{z}$ and $n$. For the data shown in Fig. 1(a), the values of $n = 0.29$, $\bar{s} = 4.07$ and $\bar{z} = 1.686$ for the colloidal gel system lead to $\kappa = (1-n)/\bar{s} = 0.17$. Remarkably, the predicted value of $\kappa$ has been observed to be very close to the experimentally obtained value of $\kappa$ for both the systems, the colloidal gel as well as the PVOH solution system. It should be emphasised that the agreement between the predicted and experimental value of $\kappa$ implies symmetric divergence of relaxation time in the vicinity of the critical gel point, which in principle has been considered as a special case. It should be noted that this work is the first report that experimentally validates the hyperscaling law for any kind of gel-forming systems.

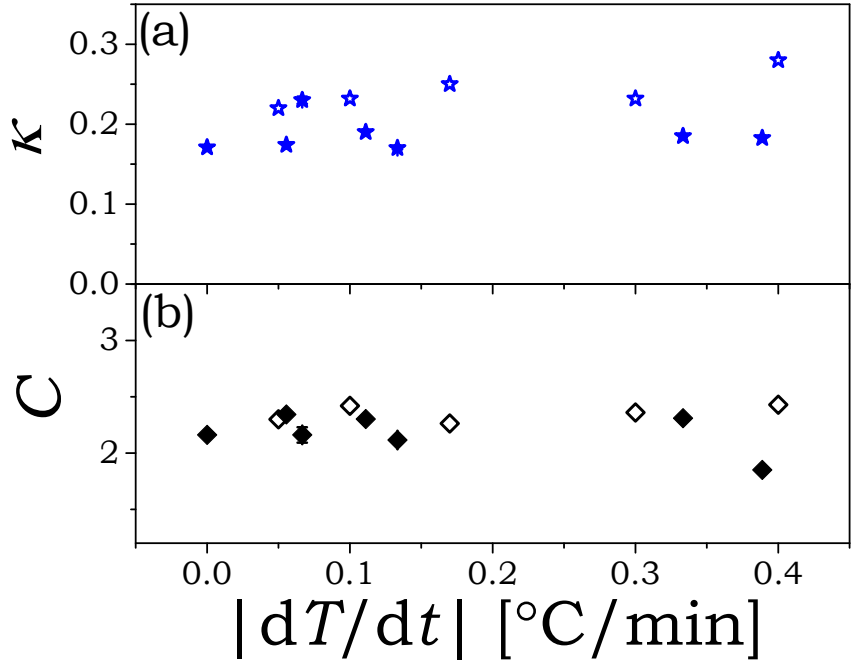

**FIG. 6.** Variation in the dynamic critical exponent $\left(\kappa\right)$ and constant of proportionality $\left(C\right)$ is plotted as a function of the rate of change of temperature for colloidal dispersion (closed symbols) and PVOH solution (open symbols). The value of $\kappa$ has been obtained from the evolution of complex modulus using Eq. (19).

The two systems studied in the present work, the colloidal dispersion of hectorite and PVOH solution have an important difference. While the gelation transition in case of former is spontaneous at any temperature, the gelation transition in the latter occurs only when the temperature is decreased. For the hectorite dispersion, as discussed before, the gelation process gets accelerated at higher temperatures. In order to check whether the above mentioned hyperscaling laws are applicable to the hectorite clay dispersion under the non-isothermal conditions, we perform oscillatory shear experiment on colloidal clay dispersion solution at different rates $\left(dT/dt\right)$ between 0.05 to 0.4°C/min starting at 10°C. In addition, we also carry out decreasing temperature ramp experiments on PVOH solution at different rates over a range -0.05 to -0.4°C/min starting at 30°C. We obtain the two parameters $\kappa$ and $C$ from the rate of change of dynamic moduli at the critical gel state. The corresponding values are plotted as a function of the magnitude of $\left|dT/dt\right|$ in Fig. 6(a) and (b). The solid symbols represent the exponents pertaining to the colloidal hectorite dispersion while the open symbols refer to the PVOH solution. It can be seen that $\kappa$ calculated through $G^*$ using Eq. (19) is always bounded between 0.17 and 0.28. Furthermore, as shown in Fig. 6(b), for all the ramp rates explored in this work on both the systems, $C$ is observed to be around 2. A value of $C=2$ means that at the gel point the growth rate of $G'$ is typically twice as that of $G''$. Interestingly, the computed value of $\kappa$ and $C$ are similar to the critical constants ($\kappa=0.2$ and $C=2$) reported for various polymeric gels [8,30,69], thus highlighting a similarity between the process of gelation irrespective of the nature of a system and the route to achieve the critical gel transition.

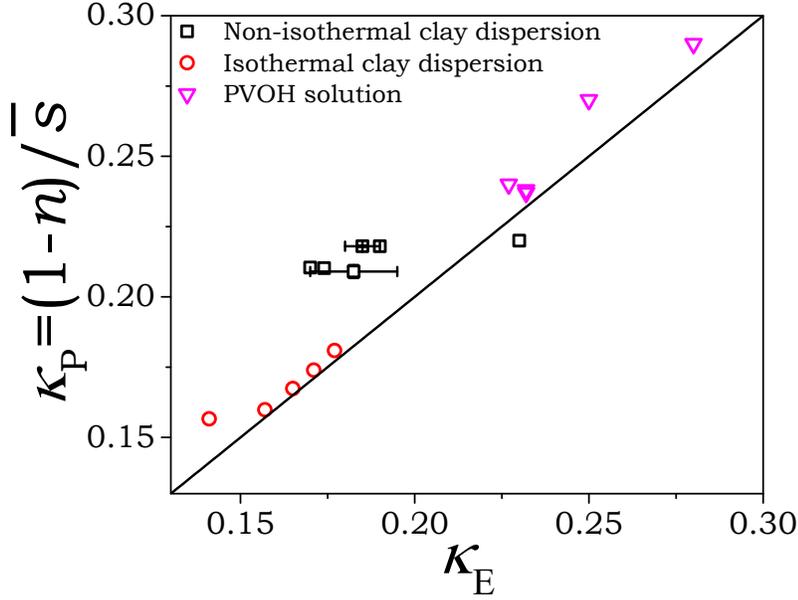

**FIG. 7.** The theoretical prediction of $\kappa$ (obtained by hyperscaling law) is plotted as a function of the experimental value of $\kappa$ (obtained from the growth rate of $G^*$) for all the systems explored in this work.

In addition to computing $\kappa$, we also compute all the critical exponents for various isothermal as well as non-isothermal experiments performed in this work on both the systems. In order to check whether the hyperscaling law (Eq. (15)) gets validated in these different experiments, we plot the value of $\kappa$ computed through $\kappa_P = (1-n)/\bar{s}$ (where subscript $P$ stands for predicted) with respect to $\kappa$ obtained experimentally using Eq. (19) (referred as $\kappa_E$, where subscript $E$ stands for experimental) in Fig. 7. The solid line denotes $\kappa_P = \kappa_E$. Remarkably, the entire data of $\kappa$ for both the systems and for isothermal as well as non-isothermal conditions is bounded in the range of 0.15-0.28 and falls in close proximity of the solid line. This leads to a very important observation that the predicted values of $\kappa$, within experimental uncertainty, are practically identical to the computed values from the growth of moduli. It should be further stated that the agreement between the predicted and experimental value of $\kappa$ corroborates the assumption of symmetric evolution of relaxation time in the vicinity of critical gel point under all the explored routes to achieve the critical gel state and progress of gelation in the post-gel regime.

This work, therefore, very clearly shows that the originally proposed critical exponent relationships work very well when $\Delta \tilde{p}$ is replaced by $\varepsilon$ that is defined either in terms of time or temperature. Interestingly the scaling laws are also fulfilled when otherwise spontaneously and isothermally aging system such as hectorite dispersion is subjected to temperature ramp, but $\varepsilon$ defined only in terms of time, despite the fact that rate of gelation is faster at higher temperatures. On the other hand, for a PVOH solution that undergoes sol-gel transition by decreasing temperature, the scaling laws get validated for different rates of cooling by representing $\varepsilon$ only in terms of temperature. Furthermore, both the systems show symmetric evolution of the relaxation time on either side of the gel state, when $\Delta \tilde{p}$ is replaced by $\varepsilon$ for all the isothermal as well as the ramp experiments. Such symmetry is in principle not obvious, and although implied implicitly through hyperscaling, this work happens to be the first one to experimentally report the same. We feel that such astonishing similarity necessitates further theoretical investigation that will render profound insights into the sol–gel transition. Moreover, the physical significance behind the growth of elastic modulus being around twice as fast as the viscous modulus at the critical gel state as observed for most of the gel-forming system needs better understanding.

The estimation of critical exponents has been carried out for over twenty experiments that include two distinctly different gel-forming systems and the experimental procedures (variation of temperature and cooling rates). These experiments suggest that the critical exponents such as $n$, $\bar{s}$, $\bar{z}$, $\bar{\nu}_S$, and $\bar{\nu}_G$, are not the universal constants but strongly depend on the microstructure and the route to achieve the critical point. The value of $n$ strictly depends on the nature of the fractal network associated with the critical gel state and is bounded between 0 and 1. The theoretical limiting cases of $n \to 0$ and $n \to 1$ respectively represents an elastic gel and viscous gel such that gel strength $S$ takes the dimension of modulus for the former and that of viscosity for the latter. In the previous publication by our group [37], we explored a wide range of concentration of synthetic hectorite clay with NaCl and observed $n$ to lie in the range of $0.15 < n < 0.6$. On the other hand, in the present work, the PVOH hydrogel $n$ has been observed to take as much large value as 0.8. Furthermore, along with $n$, $\bar{s}$ and $\bar{z}$ are also observed to vary with temperature as evident from Fig. 5. Very interestingly, while $n$, $\bar{s}$ and $\bar{z}$ have been observed to vary depending upon the system and protocol, variation of $\kappa$ over around 20 experiments carried out in the present work seems to be limited to a very narrow range of $0.196 \pm 0.01$. On considering the value of $\kappa$ obtained by Jatav and Joshi [37] over 39 systems of various concentrations of synthetic hectorite clay along with the 20 experiments of the present work, the mean value of $\kappa$ is computed to be $0.2 \pm 0.008$. Therefore, although there is no clear indication in the literature for $\kappa$ being a universal constant for critical gelation [8,53], total 59 experiments performed by our group on hectorite clay dispersion and PVOH solution do suggest value of $\kappa$ to fall in a close range near 0.2. Astonishingly, Scanlan and Winter [8] report that for over 30 chemical crosslinking experiments carried out on PDMS with different stoichiometry, chain lengths and concentrations, the value of $\kappa$ has been observed to be 0.2. This suggests $\kappa$ to have a robust value in the vicinity of 0.2 irrespective of the system and nature of gelation.

Considering a theoretical range of $n$ between $0 < n < 1$, the interrelation between $n$, $\bar{s}$, $\bar{z}$ and $\kappa$ given by Eq. (13) and (16) restricts $\bar{s}$ and $\bar{z}$ within $1/\kappa$ and 0 such that:

$$\bar{s} + \bar{z} = 1/\kappa. \tag{24}$$

This suggests that $\bar{s}$ and $\bar{z}$ are not independent to take any value bounded $1/\kappa$ and 0, but their sum must be equal to $1/\kappa$ (if we take $\kappa \approx 0.2$ as observed experimentally, we get $\bar{s} + \bar{z} \approx 5$). Furthermore, since the summation of $\bar{s}$ and $\bar{z}$ is a conserved quantity, it therefore, explains the decrease in $\bar{z}$ with the increase in $\bar{s}$ as shown in Fig. 5(b). Consequently, for a given fractal gel system, the scaling exponents are neither universal constants nor independent of each other. If $\kappa$ is considered to be a constant or bounded by a narrow range, the knowledge of $n$ fixes $\bar{s}$ and $\bar{z}$. Interestingly, the scaling exponents have also been derived theoretically. de Gennes [19] obtained $s = 0.7$ and $z = 1.7$ by drawing an analogy between phase transition and divergence of mechanical properties during the process of gelation, which leads to $n \approx 0.71$ and $\kappa \approx 0.42$. Martin *et al.* [70] estimated $s = 1.33$ and $z = 2.66$ using the percolation theory that gives $n \approx 0.67$ and $\kappa \approx 0.25$. The experimental systems, on the other hand, show $n$ to vary over a broad range signifying a very rich array of fractal microstructures associated with the critical state. Furthermore, the value of $\kappa$ obtained from de Gennes's analysis is very different than the experimentally observed narrow range around 0.2. This work, therefore also suggests that the experimentally observed value of kappa narrowly bounded around 0.2 for gel-forming systems with different gelation mechanism and microstructure demands a further theoretical investigation.

We believe that the present work is unique on multiple fronts. On one hand, our experiments on a spontaneously evolving colloidal gel and a thermo-responsive polymer gel obey all the scaling laws associated with $\eta_0, G_e$ and $\tau_{max}$ proposed in the literature. On the other hand, this work provides the quantitative measurement of symmetric divergence of dominant relaxation time in the vicinity of the critical gel point. The validation of all the scaling laws affirms the usage of the reduced parameter $\varepsilon$ for the physical gelation, which is analogous to $\Delta \tilde{p}$ for the chemical gelation process. In fact, this work is the first report on direct measurement of all the rheological critical exponents: $n$, $\bar{s}$, $\bar{z}$, $\bar{\nu}_S$, $\bar{\nu}_G$ and $\kappa$ as well as proportionality factor $C$ amongst the entire class of gel-forming systems. Furthermore, we also validate for the first time the scaling laws under isothermal and non-isothermal conditions. Most importantly, we get an excellent match between the prediction and experimental value of $\kappa$ under all the explored routes to sol-gel transition in both the systems. In a nutshell, this work for the first time carries out systematic verification of all the scaling and hyperscaling laws by a single gel-forming system among all polymeric as well as non-polymeric gels.

Table 1: List of critical scaling exponents determined experimentally at the critical gel point.

| Serial Number | System | $n$ | $s$ | $z$ | $\nu_S, \nu_G$ | $\kappa, C$ | Hyperscaling law verified |
|---|---|---|---|---|---|---|---|
| 1. | 3 wt.% LAPONITE® + 2mM NaCl (This work) | 0.29 | 4.07 | 1.686 | 5.75, 5.75 | 0.17, 2.08 | Eqs. (15) and (16) |
| 2. | 10 wt.% PVOH (This work) | 0.76 | 1.1 | 3.2 | 4.3, 4.3 | 0.22, 2.3 | Eqs. (15) and (16) |
| 3. | PDMS [8] | 0.2 to 0.9 | | | | 0.21, 2 | |
| 4. | Tetraethoxysilane [10] | 0.7 to 0.72 | 0.8 to 1 | 2.2 to 2.6 | | | Eq. (15) |
| 5. | Polyurethane [11,71] | | 0.6 to 0.9 | 1.8 to 3.9 | | | |
| 6. | Pectin [34] | 0.7 | 0.82 | 1.93 | | | Eq. (15) |
| 7. | Randomly branched polyster [25] | 0.66 | 1.36 | 2.71 | | | Eq. (15) |
| 8. | Gelatin [26] | 0.35 to 0.6 | 1 to 2.7 | 1.5 to 2.5 | | | Eq. (15) |
| 9. | Alginate [72] | 0.5 to 0.65 | 1.3 to 1.6 | 1.5 to 2.8 | | | Eq. (15) |
| 10. | Polyacrylonitrile [33] | 0.65 | 0.68 | 1.6 | | | Eq. (15) |
| 11. | Cellulose [73] | 0.88 | 0.54 | 3.95 | | | Eq. (15) |
| 12. | Chitosan-$\beta$-glycerophosphate [27] | 0.5 to 0.66 | 0.74 to 1 | 1.4 to 2.3 | | | Eq. (15) |
| 13. | Gellan gum [28] | 0.38 to 0.77 | 0.5 to 2.2 | 1.4 to 1.7 | | | Eq. (15) |
| 14. | Peptide hydrogels [21] | 0.6 | | 2.9 | 4.8, 4.8 | | Eq. (16) |
| 15. | Colloidal silica particles with octadecyl chains in decalin [53] | 0.5 | | | | 0.25, 2 | |
| 16. | 1.4 to 4 wt.% LAPONITE® + 0 to 7 mM NaCl [37] | 0.15 to 0.6 | | | | 0.1 to 0.3, 2 to 5 | |
| 17. | $\kappa$ – carrageenan [29] | 0.62 | 1.6 | 2.7 | | | Eq. (15) |

### IV. Concluding remarks

In this work, we investigate the application of the scaling theories on two diverse systems that undergo physical sol-gel transition. While the aqueous dispersion of synthetic hectorite evolves spontaneously to form interparticle bonds while undergoing the sol-gel transition, PVOH solution undergoes physical gelation upon cooling. On performing oscillatory shear experiments, both the systems demonstrate all the characteristic rheological signatures of critical gel transition, which suggests the presence of the weakest space spanning percolated network. The dynamic measurements allow experimental evaluation of the zero-shear viscosity of the pre gel state and the equilibrium modulus of the post gel state. Below the critical point, $\eta_0$ is observed to vary as $\eta_0 \sim \varepsilon^{-\bar{s}}$ with the relative distance, expressed analogous to chemical gelation, as: $\varepsilon = |t_c - t|/t_c$ for the colloidal gel and $\varepsilon = |T_c - T|/T_c$ for the thermo-responsive polymer gel. The subsequent growth of $G_e$ is observed to follow $G_e \sim \varepsilon^{\bar{z}}$ relationship in the post-critical gel regime. Very interestingly, the estimated dominant relaxation timescales of both the systems show a symmetric divergence of $\tau_{max}$ on either side of the critical gel state. Remarkably, the critical relaxation exponent computed using the scaling exponents are observed to be identical to the value of $n$ obtained from the Winter criteria. Furthermore, to test the applicability of the scaling theories under non-isothermal condition, similar oscillatory experiments are performed at different temperature ramp rates on both the physical gels. Remarkably, all the scaling laws have been found to be obeyed not only under isothermal conditions but also for non-isothermal conditions. Moreover, the predicted dynamic critical exponent $\kappa$ from the scaling theories agreed with the one calculated using Winter criteria very well. Many experiments performed on both the systems with varying routes to achieve the critical gel state indicate that $\kappa$ to fall in a close range in the vicinity $\kappa = 0.2$. However, the scaling exponents $n$, $\bar{s}$, $\bar{z}$, $\bar{\nu}_S$ and $\bar{\nu}_G$ are dependent on the microstructure of the critical gel state and are interrelated by the hyperscaling laws. We observe that despite the significantly differing mechanism of gelation in the colloidal gel and polymer solution as well as the fractal dimension, the critical gel state obeyed all the scaling and hyperscaling laws.

**Acknowledgement:** We acknowledge the financial support from the Science and Engineering Research Board (SERB), Department of Science and Technology,

Government of India. We thank Prof. Henning Winter for discussion and for insightful comments on this work.